\documentclass[a4paper,11pt]{article}
\usepackage{pos}

\title{Precision predictions for scalar leptoquark pair production at the LHC}

\author*[a, \dag]{Christoph Borschensky}
\author[b]{Benjamin Fuks}
\author[c]{Anna Kulesza}
\author[c, \ddag]{Daniel Schwartl\"ander}

\affiliation[a]{Institute for Theoretical Physics, University of T\"ubingen, Auf der Morgenstelle 14, 72076 T\"ubingen, Germany}
\affiliation[b]{Laboratoire de Physique Th\'eorique et Hautes Energies (LPTHE), UMR 7589, Sorbonne Universit\'e et CNRS, 4 place Jussieu, 75252 Paris Cedex 05, France}
\affiliation[c]{Institute for Theoretical Physics, WWU M\"unster, D-48149 M\"unster, Germany}

\note[\dag]{We acknowledge support by the state of Baden-W\"urttemberg through bwHPC and the DFG Grant No.\ INST 39/963-1 FUGG (bwForCluster NEMO). Numerical calculations have been also performed on the PALMA HPC cluster of the WWU M\"unster.}
\note[\ddag]{partially supported by the DFG Grant No. KU3103/2}

\emailAdd{christoph.borschensky@uni-tuebingen.de}
\emailAdd{fuks@lpthe.jussieu.fr}
\emailAdd{anna.kulesza@uni-muenster.de}
\emailAdd{d\_schw20@uni-muenster.de}

\abstract{We present precision predictions for scalar leptoquark pair production at the LHC. Apart from QCD contributions, included are the lepton $t$-channel exchange diagrams relevant in the light of the recent $B$-flavour anomalies. All contributions are evaluated at next-to-leading order in QCD and improved by resummation, in the threshold regime, of the corrections from soft-gluon radiation at the next-to-next-to-leading-logarithmic accuracy. All corrections are found equally relevant. Furthermore, the impact of different sets of parton distribution functions is discussed. These predictions constitute the most precise leptoquark cross section calculations available to date and are necessary for the best exploitation of leptoquark LHC searches.}

\FullConference{%
  *** The European Physical Society Conference on High Energy Physics (EPS-HEP2021), ***\\
  *** 26-30 July 2021 ***\\
  *** Online conference, jointly organized by Universität Hamburg and the research center DESY ***
}

\newcommand{\alphas}{\alpha_{\text{s}}}
\newcommand{\lqp}{{\text{LQ}\,\text{LQ}^*}}
\newcommand{\mlq}{m_{\text{LQ}}}
\newcommand{\pwb}{{\textsc{POWHEG-BOX}}}

\newcommand{\mgamc}{{\textsc{MadGraph5\_aMC@NLO}}}

\begin{document}
\maketitle

\section{Introduction}
Scalar leptoquarks are bosonic particles beyond the Standard Model which couple to both quarks and leptons via a Yukawa-type interaction, and which were originally proposed in the context of Grand Unification. Over the recent years, the appearance of so-called flavour anomalies, namely discrepancies between theoretical expectations and experimental measurements for certain flavour observables such as the $R_{K^{(*)}}$ and $R_{D^{(*)}}$ ratios pertaining to lepton-flavour universality (see e.g.\ \cite{Lees:2012xj,Belle:2019rba,Aaij:2019wad}), has led to increased interest in leptoquark models. These are known to mitigate or even resolve the tensions. Until now, collider experiments such as the Large Hadron Collider (LHC) have not seen any signals of leptoquark production and the current exclusion limits require leptoquark masses to be larger than about 1.0--1.8 TeV, depending on the specifics of the model, see e.g.\ \cite{Aad:2020iuy,CMS:2020wzx}.

\begin{figure}[t]
	\setlength\tabcolsep{0pt}
	\centering
	\begin{tabular}{cccc}
		\includegraphics[width=.25\textwidth]{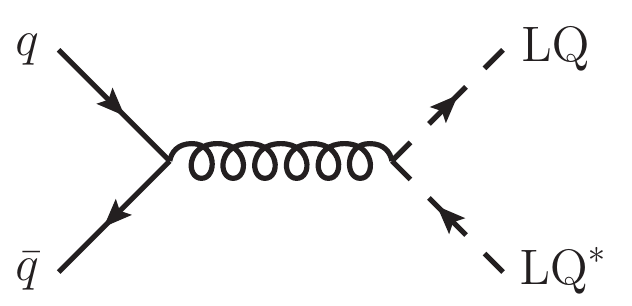} & \includegraphics[width=.25\textwidth]{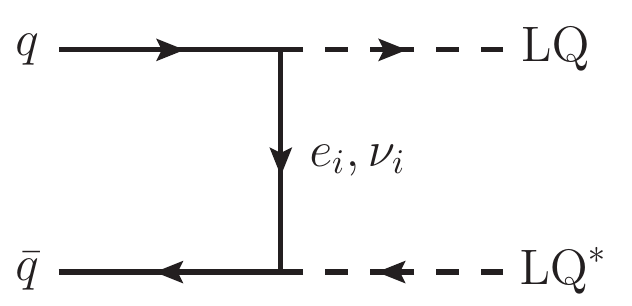} & \includegraphics[width=.25\textwidth]{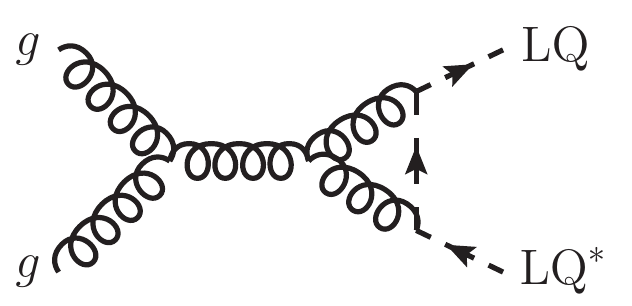} & \includegraphics[width=.25\textwidth]{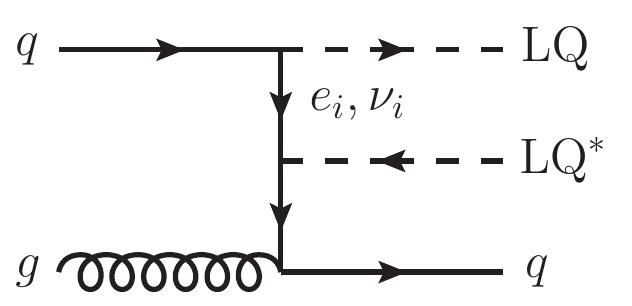}\\
		(a) & (b) & (c) & (d)
	\end{tabular}
	\caption{Representative Feynman diagrams for scalar leptoquark pair production, with pure-QCD (a) and leptonic $t$-channel (b) contributions at tree level, and examples for virtual (c) and real (d) QCD corrections.}
	\label{th:feyndiags}
\end{figure}
Previous direct search studies of leptoquark pair production typically neglected contributions proportional to the leptoquark-lepton-quark Yukawa couplings in relation to the leading pure-QCD terms, cf.\ figure~\ref{th:feyndiags}~(a). Explanations to the flavour anomalies however require Yukawa couplings of $\mathcal{O}(1)$ and masses of $\mathcal{O}(\text{TeV})$. The inclusion of leptonic $t$-channel contributions of figure~\ref{th:feyndiags}~(b) as well as QCD corrections up to the next-to-leading order (NLO) in the strong coupling $\alphas$, cf.\ figure~\ref{th:feyndiags}~(c) and (d), and threshold resummation corrections could thus impact the predictions notably, as shown in our recent works \cite{Borschensky:2020hot,Borschensky:2021hbo}. In these proceedings, we discuss the most important results.

\section{Theoretical setup}
In our simplified framework, we extend the Standard Model (SM) by five species of scalar leptoquarks that couple to quarks and leptons, following standard notation~\cite{Buchmuller:1986zs}: $S_1$, $\tilde S_1$, $R_2$, $\tilde R_2$, and $S_3$. They lie in the $(\mathbf{3}, \mathbf{1})_{-1/3}$, $(\mathbf{3}, \mathbf{1})_{-4/3}$, $(\mathbf{3}, \mathbf{2})_{7/6}$, $(\mathbf{3}, \mathbf{2})_{1/6}$, and $(\mathbf{3}, \mathbf{3})_{-1/3}$ representations of the SM gauge group $SU(3)_C \times SU(2)_L \times U(1)_Y$, respectively, where the bold numbers denote the transformation properties with respect to the $SU(3)_C$ and $SU(2)_L$ gauge groups, and the subscript indicates the hypercharge. Then, the Lagrangian describing the leptoquark interactions is:
\begin{equation}
	\begin{split}
		\mathcal{L}_{\mathrm{LQ}} = \mathcal{L}_{\mathrm{kin.}} &\ + \mathbf{y_1^{RR}} \bar u_R^c e_R S_1^\dag + \mathbf{y_1^{LL}} \left(\bar Q_L^c \cdot L_L\right) S_1^\dag + \mathbf{\tilde y_1^{RR}} \bar d_R^c e_R \tilde S_1^\dag + \mathbf{y_2^{LR}} \bar e_R Q_L R_2^\dag\\
		&\ + \mathbf{y_2^{RL}} \bar u_R \big(L_L \cdot R_2\big) + \mathbf{\tilde y_2^{RL}} \bar d_R \big(L_L \cdot \tilde R_2\big) + \mathbf{y_3^{LL}} \big(\bar Q_L^c \cdot \sigma_k L_L\big) \big(S_3^k\big)^\dag + \mathrm{H.c.},
	\end{split}
	\label{eq:lag}
\end{equation}
where $\mathcal{L}_{\mathrm{kin.}}$ collects all gauge-invariant kinetic and mass terms and the Yukawa couplings $\mathbf{y}/\mathbf{\tilde y}$ are $3\times 3$ matrices in flavour space, the first (second) index of any element $y_{ij}$ referring to the quark (lepton) generation. We generically denote the leptoquark mass by $\mlq$. In terms of their component fields with a specific electric charge, the electroweak multiplets can be written, with the matrix representation of the triplet $S_3 = 1/\sqrt 2\,\sigma_k S_3^k$ and the Pauli matrices $\sigma_k$ for $k = 1, 2, 3$, as:
\begin{equation}
	\setlength{\arraycolsep}{1pt}
	\renewcommand{\arraystretch}{1.3}
	\begin{split}
		S_1 = S_1^{(-1/3)},\ \tilde S_1 = \tilde S_1^{(-4/3)},\ R_2 = \begin{pmatrix}R_2^{(+5/3)}\\R_2^{(+2/3)}\end{pmatrix},\ \tilde R_2 = \begin{pmatrix}\tilde R_2^{(+2/3)}\\\tilde R_2^{(-1/3)}\end{pmatrix},\ S_3 =  \begin{pmatrix}\frac{1}{\sqrt{2}}S_3^{(-1/3)} & S_3^{(+2/3)}\\S_3^{(-4/3)} & -\frac{1}{\sqrt{2}}S_3^{(-1/3)}\end{pmatrix}.
	\end{split}
\end{equation}
In our studies, we consider a simplified scenario in which the Standard Model is extended by either only the $S_1$ or the $R_2$ species, as well as three benchmark scenarios motivated by a simultaneous resolution of the $R_{K^{(*)}}$ and $R_{D^{(*)}}$ anomalies:
\emph{(\textbf{a})} a solution involving only $R_2$, and two-leptoquark explanations with either \emph{(\textbf{b})} both $R_2$ and $S_3$ or \emph{(\textbf{c})} both $S_1$ and $S_3$ (see section 2.1.2 of \cite{Borschensky:2021hbo}).

We calculate the fixed-order cross section including NLO-QCD corrections for the pair production of scalar leptoquarks at the LHC. Our results consistently include all contributions from figure~\ref{th:feyndiags}, i.e.\ the squares of pure-QCD and $t$-channel diagrams as well as their interference with terms of $\mathcal{O}(\alphas^2,\, y^4,\, y^2\alphas)$ at Born level and $\mathcal{O}(\alphas^3,\, y^4 \alphas,\, y^2 \alphas^2)$ for the QCD corrections, respectively. We consider the sum of all three classes of terms as our complete NLO-accurate prediction. The results are implemented in the \mgamc{}~\cite{Alwall:2014hca} and \pwb{}~\cite{Nason:2004rx} frameworks.

Moreover, we consider corrections from the emission of soft gluons in the threshold limit $\beta^2 = 1 - 4\mlq^2/s \to 0$ with the partonic centre-of-mass energy $s$ by resumming logarithms $\alphas^n \ln^k \beta^2$ with $k \le 2n$ to all orders. We apply the Mellin-space formalism to write the resummed cross section, now depending on the Mellin-moment $N$, in the factorised form \cite{Sterman:1986aj}:\vspace{-.9mm}
\begin{equation}
	\tilde \sigma^{\mathrm{res, NNLL}}_{ij\rightarrow \lqp,I}(N) = \tilde\sigma^{(0)}_{ij\rightarrow \lqp,I}(N)\,\tilde C_{ij\rightarrow \lqp,I}(N)\,\Delta^{S}_I(N+1)\,\Delta_i(N+1)\,\Delta_j(N+1),
\end{equation}
with $I = \mathbf{1}\text{ (singlet)}, \mathbf{8}\text{ (octet)}$ indicating the colour representation of the final state. The Mellin-transformed Born cross section is $\tilde\sigma^{(0)}_{ij\rightarrow \lqp,I}$, the hard-matching coefficients $\tilde C_{ij\rightarrow \lqp,I}$ collect non-logarithmic higher-order terms, and the functions $\Delta^{S}_I\Delta_i\Delta_j$ contain the resummed soft-collinear logarithms. The result is then matched to the fixed-order calculation to avoid double-counting, and transformed back to physical momentum space via an inverse Mellin transform. Here, we consider threshold resummation up to next-to-next-to-leading-logarithmic (NNLL) accuracy.

\section{Precision predictions}
We denote our prediction including $t$-channel and resummation corrections as ``NLO w/ $t$-channel + NNLL''. The results are compared to pure-QCD predictions labeled ``NLO-QCD''. All calculations are carried out for a centre-of-mass energy of $\sqrt{S} = 13$ TeV, employing three different sets of parton distribution functions (PDFs), namely CT18~\cite{Hou:2019efy}, NNPDF3.1~\cite{NNPDF:2017mvq}, and MSHT20~\cite{Bailey:2020ooq} for the description of the proton's parton content. The central renormalisation and factorisation scales are set to $\mu_R = \mu_F = \mlq$, and the scale uncertainty is evaluated through the 7-point method by varying the scales up and down by a factor of 2 relative to the central value.

\begin{figure}[t]
	\centering
	\includegraphics[width=0.379\textwidth]{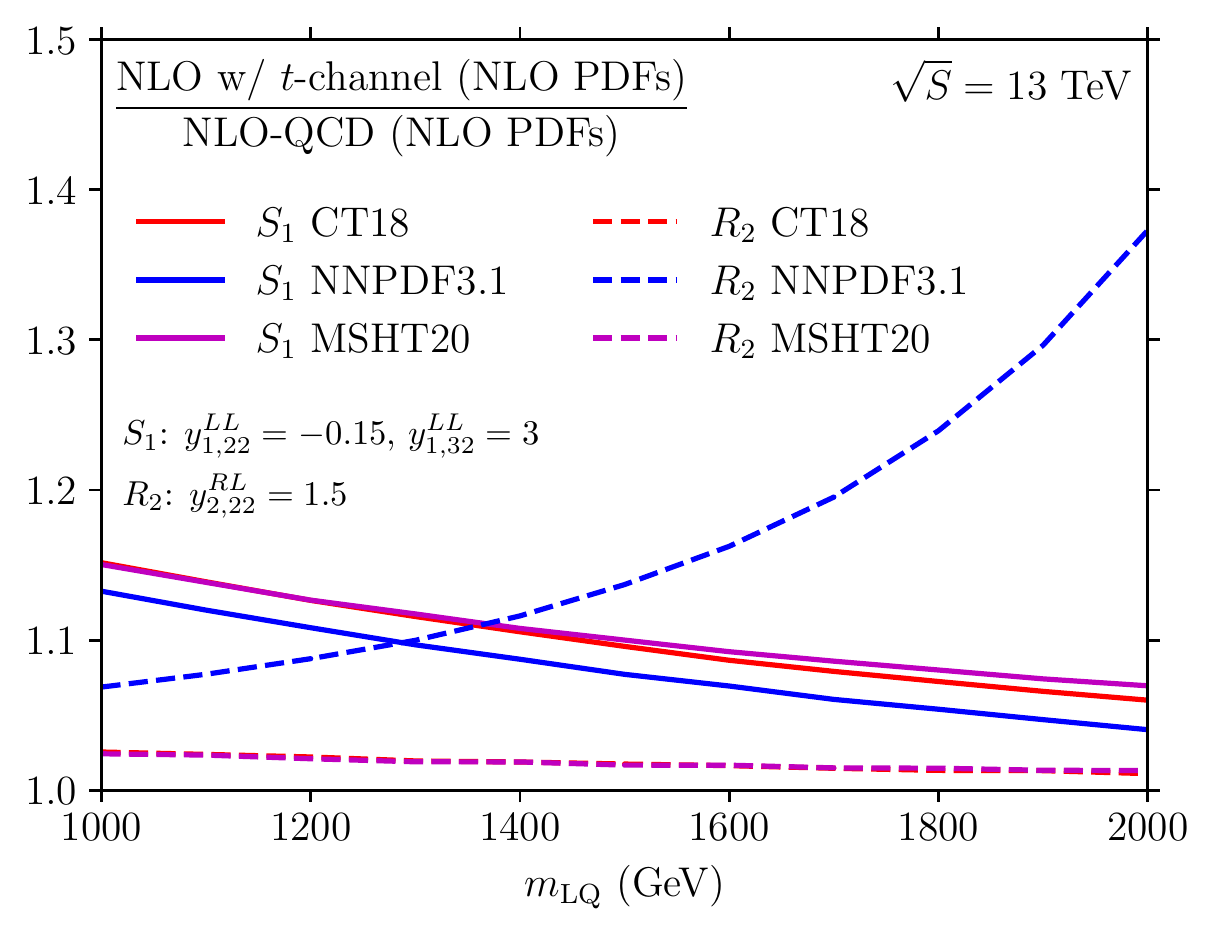}\hspace{5mm}\includegraphics[width=0.379\textwidth]{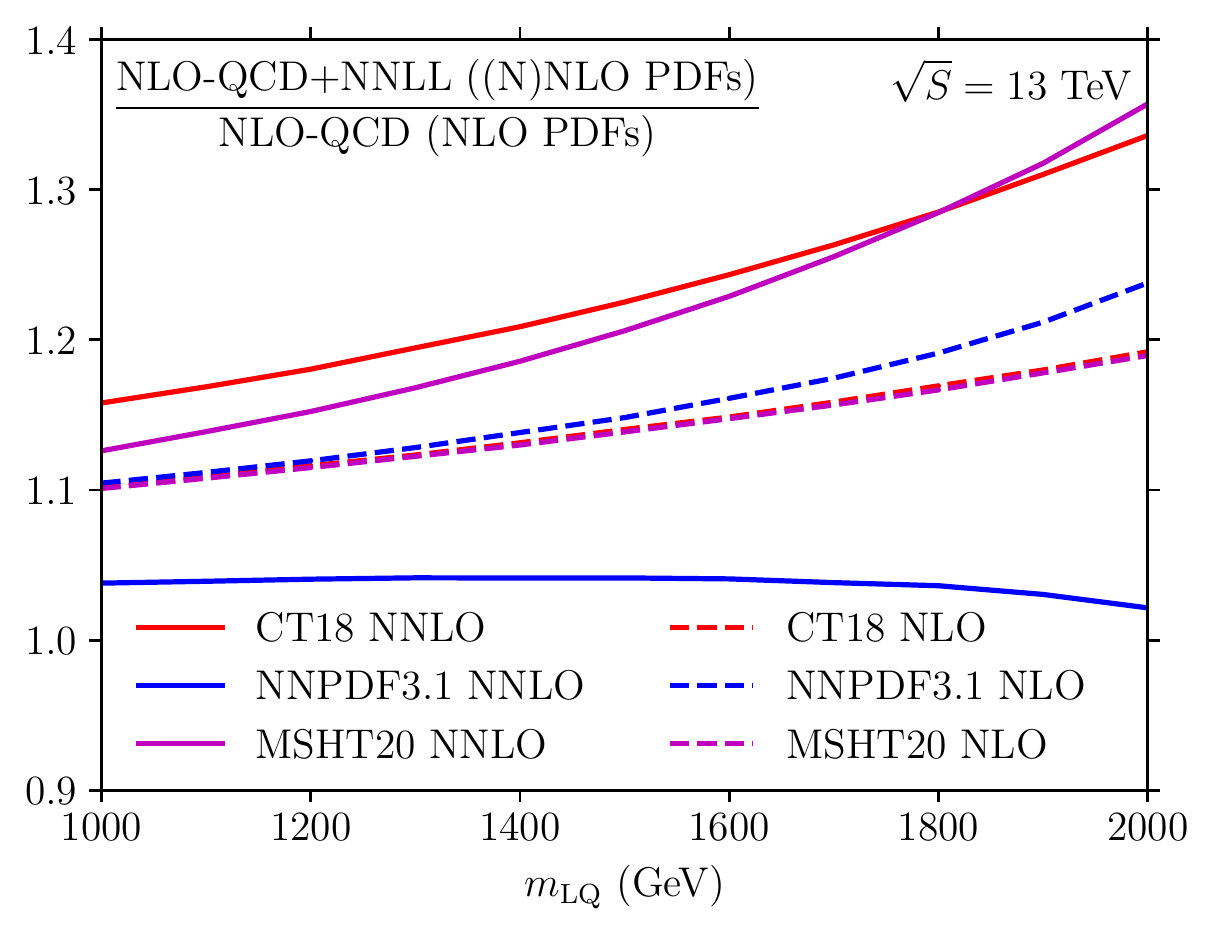}\\
	\includegraphics[width=0.379\textwidth]{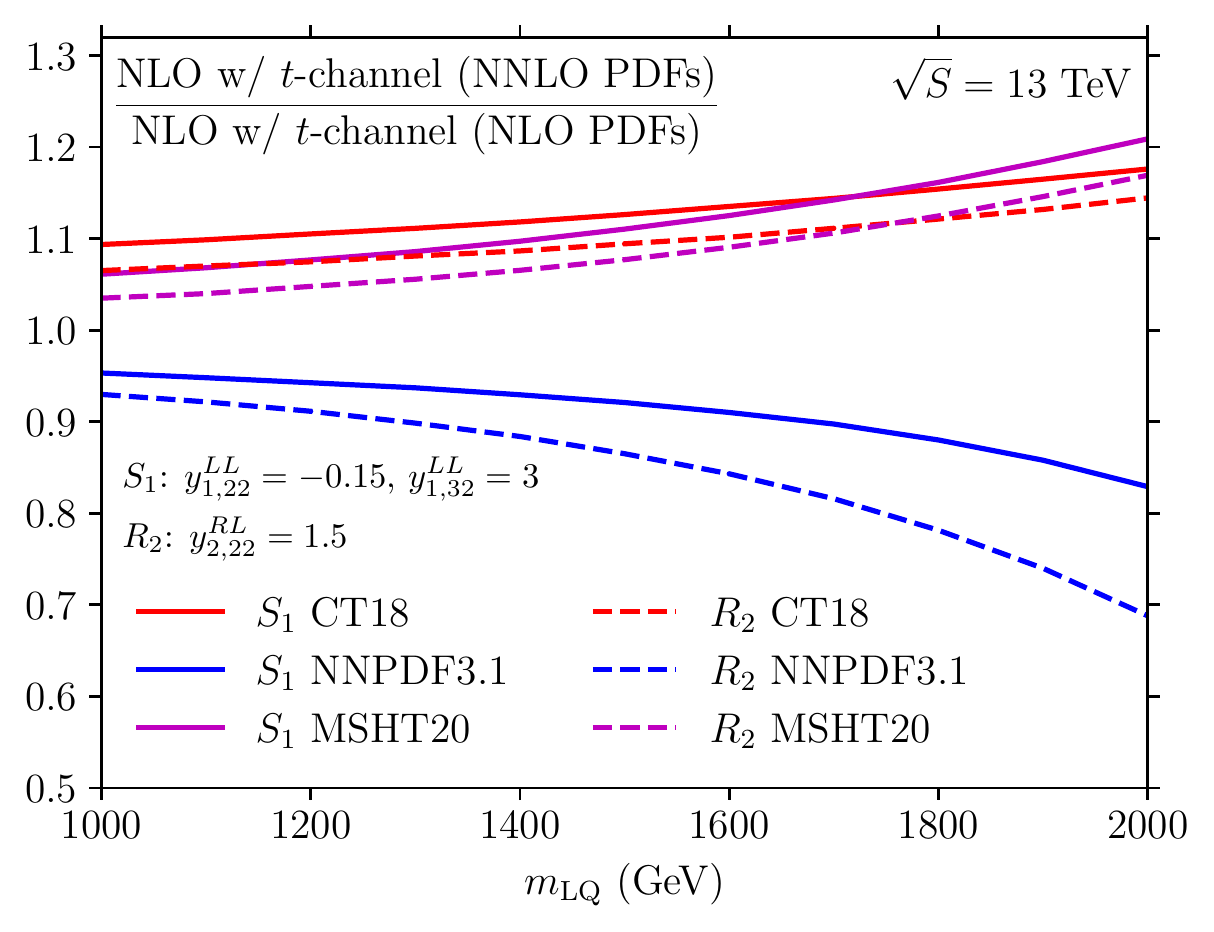}\hspace{5mm}\includegraphics[width=0.379\textwidth]{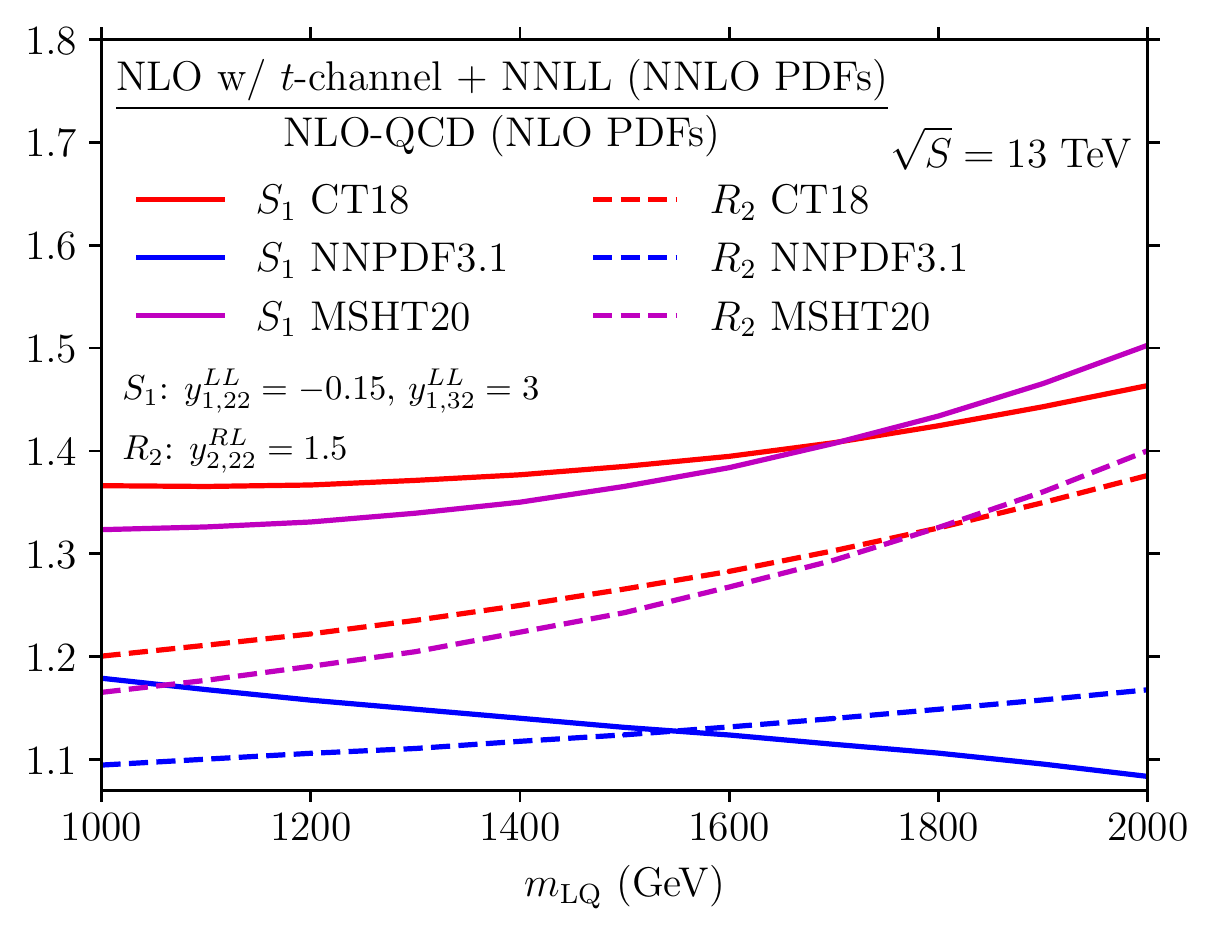}
	\caption{Impact of various contributions on the predictions associated with $S_1^{(-1/3)}$ and $R_2^{(+5/3)}$ pair production, shown as ratios. Top left: $t$-channel contributions. Top right: threshold resummation corrections (independent of the leptoquark model). Bottom left: choice of PDFs. Bottom right: combined effects.}
	\label{fig:lq:rel-importance}
\end{figure}
We begin with an analysis of the impact of the various contributions considered in this work. We assume only one leptoquark species to be present and discuss the pair production of the $S_1^{(-1/3)}$ and $R_2^{(+5/3)}$ eigenstates. In figure~\ref{fig:lq:rel-importance}, we present ratios to highlight the relative importance: NLO w/ $t$-channel over NLO-QCD to assess the impact of $t$-channel contributions (top left), NLO-QCD + NNLL over NLO-QCD to evaluate the size of the resummed corrections (top right), NLO w/ $t$-channel with NNLO PDFs over the same with NLO PDFs to analyse the PDF choice (bottom left), and NLO w/ $t$-channel + NNLL over NLO-QCD to show the combined effect of all contributions (bottom right). It can be seen that all pieces are of similar size, possibly increasing or reducing the predictions by a few tens of per cent. While the CT18 and MSHT20 predictions are generally similar with an often very different behaviour for NNPDF3.1 related to the treatment of the charm quark PDF, the effects depend strongly on the flavour structure of the leptoquark coupling. It is therefore important to consider the combination of all contributions as no generic behaviour arises.

\begin{figure}[t]
	\centering
	\includegraphics[width=.333\textwidth]{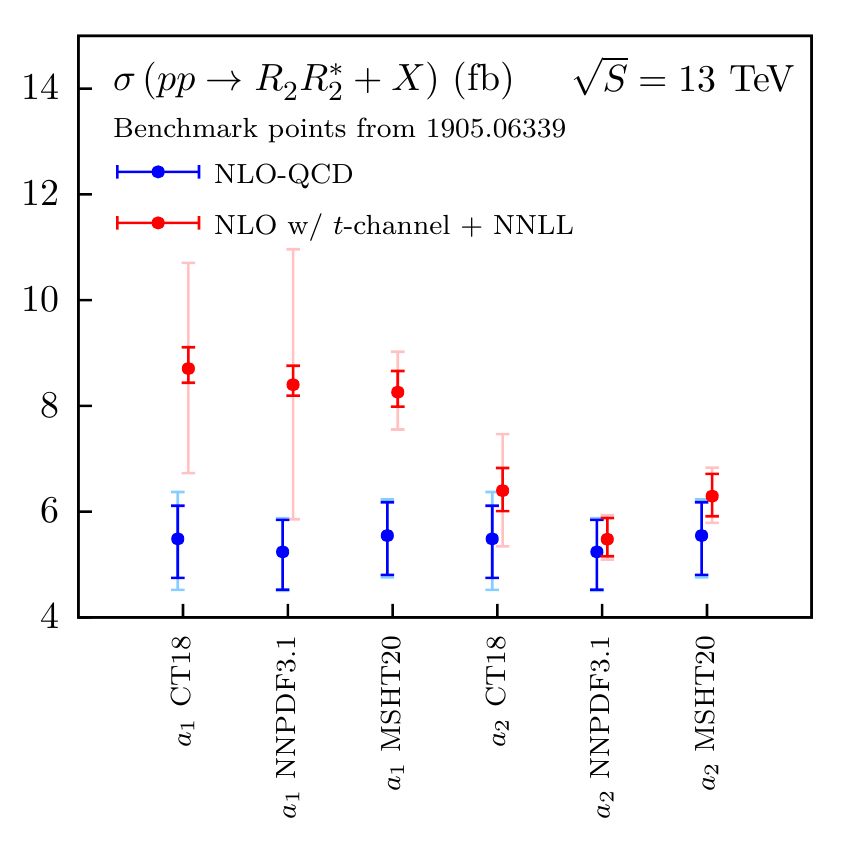}\includegraphics[width=.333\textwidth]{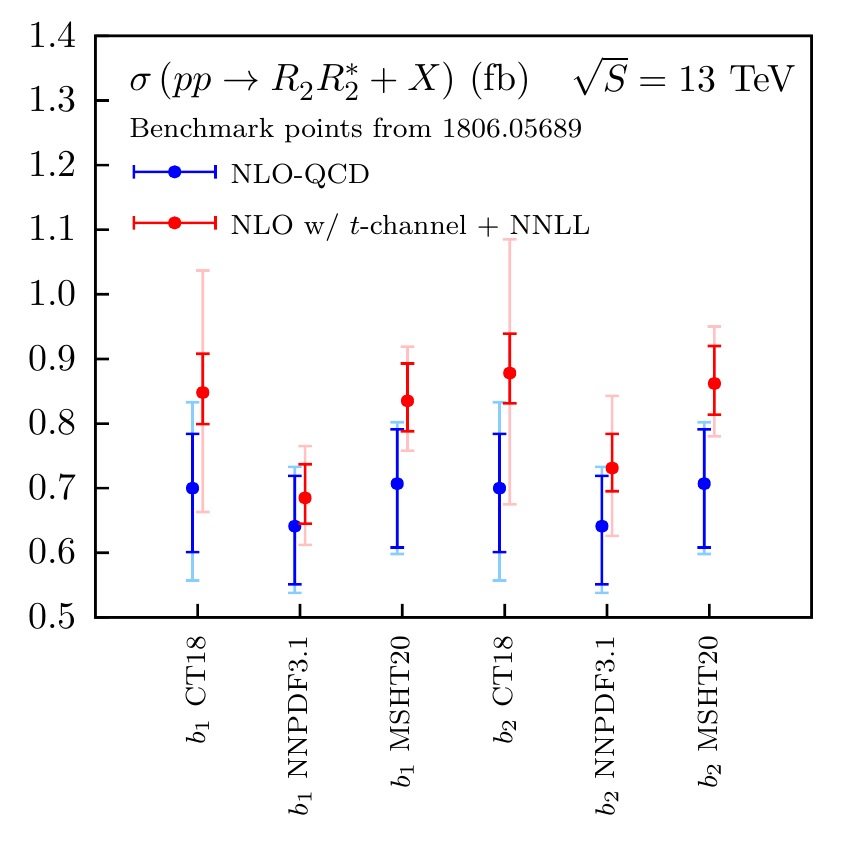}\includegraphics[width=.333\textwidth]{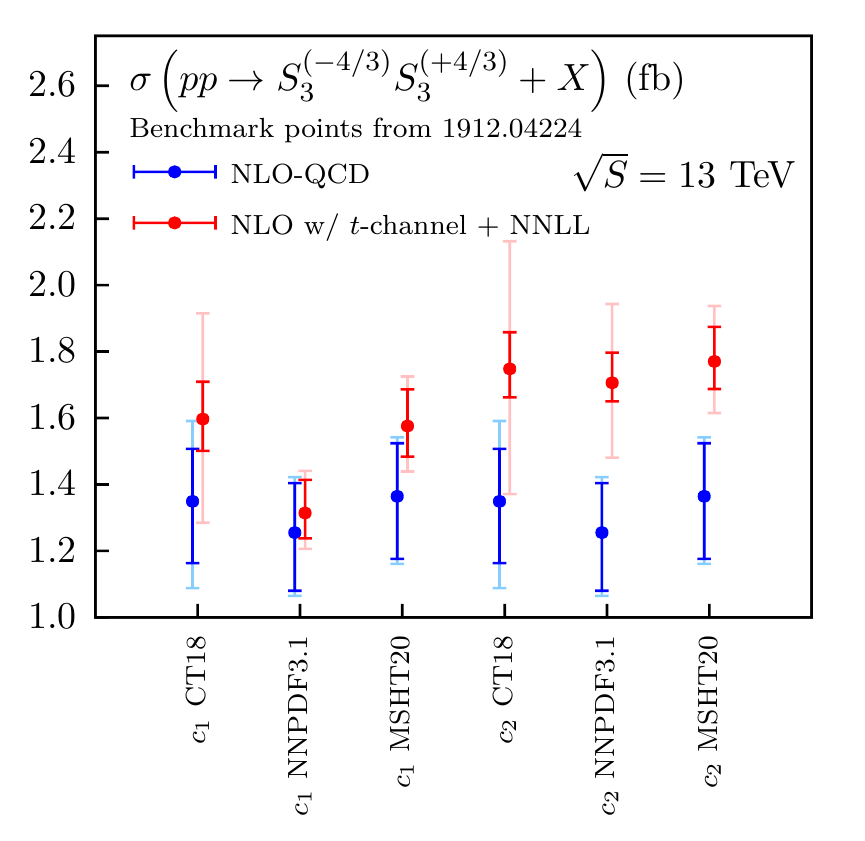}
	\caption{Comparison of total cross section predictions at NLO-QCD (blue) and NLO w/ $t$-channel + NNLL (red), for three benchmark scenarios \emph{(\textbf{a})}, \emph{(\textbf{b})}, and \emph{(\textbf{c})} (see \cite{Borschensky:2021hbo} for further information). The dark-coloured error bars denote the scale uncertainties, and the light-coloured ones their combination with the PDF uncertainties.}
	\label{fig:benchmarksplots}
\end{figure}
Next, we discuss in figure~\ref{fig:benchmarksplots} predictions for total cross sections evaluated in the three phenomenologically motivated benchmark scenarios \emph{(\textbf{a})}, \emph{(\textbf{b})}, and \emph{(\textbf{c})}, including a full error analysis with scale and PDF uncertainties. We select two points in the allowed parameter space from each benchmark, and compare NLO-QCD with the NLO w/ $t$-channel + NNLL predictions, evaluated with NLO and NNLO PDF sets, respectively. A comparison of the dark-coloured bands between the two accuracies shows that the NNLL corrections greatly improve the scale behaviour. In contrast, with the exception of MSHT20 being the most recent of the PDF sets considered, the full uncertainties grow for NLO w/ $t$-channel + NNLL which can be attributed to the difference between NLO and NNLO PDFs. While for some points, the two accuracies agree within errors, in several cases, the new contributions lead to a notable enhancement outside of the error bands, as seen mainly for $a_1$ and $c_2$ in the leftmost and rightmost plots. Thus, NLO-QCD cannot reliably approximate the full pair production process, in particular for new generations of PDFs with smaller uncertainties.

\section{Conclusions}
We have calculated precision predictions for the pair production of scalar leptoquarks at the LHC. Included are QCD and leptonic $t$-channel contributions up to NLO-QCD and threshold resummation corrections up to NNLL accuracy. Our results constitute the most precise theoretical predictions for this class of processes to date. In light of the large Yukawa couplings and leptoquark masses required for a solution to the flavour anomalies, the corrections we have considered become particularly relevant. We have observed that all classes of contributions are equally important and can impact the predictions in often contrasting ways. The developed codes and numerical tables in the NNLL-fast format are available publicly from:\par
{\centering
	\url{https://www.uni-muenster.de/Physik.TP/research/kulesza/leptoquarks.html}
\par}

\setlength{\bibsep}{0pt}

\providecommand{\href}[2]{#2}\begingroup\raggedright\endgroup

\end{document}